\documentclass[aps,pra,groupaddress,showpacs,twocolumn]{revtex4-1}
\usepackage{amsfonts}
\usepackage{bm}
\usepackage{epsfig,amsmath,graphicx,amssymb}
\usepackage{color}

\def\be{\begin{equation}}
\def\ee{\end{equation}}
\def\bee{\begin{eqnarray}}
\def\ene{\end{eqnarray}}
\def\bes{\begin{subequations}}
\def\ees{\end{subequations}}

\begin{document}

\title{Rogue waves, rational solitons, and modulational instability in an
integrable fifth-order nonlinear Schr\"{o}dinger equation}
\author{Yunqing Yang$^{1,2}$}
\author{Zhenya Yan$^{3}$}
\email{zyyan@mmrc.iss.ac.cn}
\author{Boris A. Malomed$^{4}$}

\affiliation{\vspace{0.1in} $^{1}$School of Mathematics, Physics and Information Science, Zhejiang Ocean University, Zhoushan, Zhejiang  316022, China \\
$^{2}$Key Laboratory of Oceanographic Big Data Mining $\&$ Application \\ of Zhejiang Province, Zhoushan, Zhejiang 316022, China \\
$^3$Key Laboratory of Mathematics Mechanization, Institute of Systems
Science, AMSS, Chinese Academy of Sciences, Beijing 100190, China \\
$^4$Department of Physical Electronics, School of Electrical Engineering, Tel Aviv University,
Tel Aviv 59978, Israel}

\begin{abstract}
We analytically study rogue-wave (RW) solutions and rational solitons of an integrable fifth-order
nonlinear Schr\"{o}dinger (FONLS) equation with three free parameters. It
includes, as particular cases, the usual NLS, Hirota, and
Lakshmanan-Porsezian-Daniel (LPD) equations. We present continuous-wave (CW)
solutions and conditions for their modulation instability in the framework
of this model. Applying the Darboux transformation to the CW input, novel
first- and second-order RW solutions of the FONLS equation are analytically found. In particular,
trajectories of motion of peaks and depressions of profiles of the first- and second-order RWs are
produced by means of analytical and numerical methods. The solutions also
include newly found rational and W-shaped one- and two-soliton modes. The
results predict the corresponding dynamical phenomena in extended models of
nonlinear fiber optics and other physically relevant integrable systems.
\end{abstract}

\maketitle


\baselineskip=12pt


\textbf{Solitons are usually generated by the balance of linear  dispersion
and nonlinear self-focusing. Rogue waves (RWs), as a special type of
solitary waves, are driven by a similar mechanism acting on top of a
modulationally unstable flat background (alias continuous wave, CW). In
particular, the celebrated nonlinear Schr\"{o}dinger (NLS) equation with the
self-focusing sign of the nonlinearity gives rise to both bright solitons
and RWs, which are used to model nonlinear phenomena in diverse fields, such
as nonlinear optics, deep ocean, plasmas, Bose-Einstein condensates, and
biophysics. In nonlinear fiber optics, the correct
description of tall narrow pulses produced by the RWs makes necessary to
include higher-order effects, such as the stimulated Raman scattering,
nonlinear group-velocity dispersion (alias the shock term), and third-order
dispersion. In the general case, the addition of such terms to the NLS
equation breaks its integrability, making it impossible to find solitons and
RWs in an analytical form. However, under special conditions the respective
extended NLS equation may keep the integrability. In particular, quite a
general model of the latter type, namely, the fifth-order NLS (FONLS)
equation, which includes three free parameters, was recently found. Its
particular forms amount to other previously known integrable equations, such
as the usual NLS, Hirota, and Lakshmanan-Porsezian-Daniel (LPD) equations.
In this paper, our objective is to use the integrability of the FONLS
equation for constructing its exact RW solutions. Because RWs are driven by
the modulational instability of flat CW states, we first present the CW
solutions of this model and conditions for their modulation instability.
Then we proceed to the core part of the analysis, which reveals novel exact
first- and second-order RW solutions, by means of the Darboux transformation
applied to the CW input. The solutions are obtained in the form of explicit
rational functions of the spatial coordinate and time. In particular,
interesting results are produced by the analysis of trajectories of the
motion of peaks and depressions of the RW profiles. Furthermore, the
obtained rational solutions include new W-shaped one- and two-soliton modes
of the FONLS equation. The analysis developed in this work and results
produced by it may suggest new possibilities to analyze other sophisticated
nonlinear models of practical significance.}

\section{Introduction}

Rogue waves (RWs, also known as monster waves, killer waves, giant waves,
extreme waves, etc.)~\cite{rw1, rw2}, as a kind of localized modes, are
generated, in the paradigmatic form~\cite{peregrine1983water}, by the
self-focusing nonlinear Schr\"{o}dinger (NLS) equation for wave amplitude $%
\psi \left( x,t\right) $,
\begin{equation}
i\psi _{x}+\frac{1}{2}\psi _{tt}+|\psi |^{2}\psi =0.  \label{nls}
\end{equation}%
as a limit case of the Ma-breather~\cite{ma} or Akhmediev-breather~\cite%
{nail86} solutions on top of a flat continuous-wave (CW) state, which is
subject to the modulational instability (MI)~\cite{nail86}. This equation is
written in terms of the guided-wave notation, with $x$ and $t$ being the
propagation distance and reduced time, respectively \cite{Agrawal,opt}. The RWs
are sometimes called \textit{rogons}, if they reappear unaffected in the
size and shape after their interactions~\cite{rogon}.

The RW phenomenon was initially found in the deep ocean~\cite{rw1, rw2}.
Recently, it has also drawn growing interest in other fields, such as
Bose-Einstein condensates (BEC), nonlinear optics, plasma physics, and even
dynamics of financial markets~\cite%
{Garrett2008rogue,yan2010three,solli2007rogue,becrw,chen2012optical,akhmediev2013recent,yan2011finace}%
. Generally speaking, the RWs are spatially and temporally localized
excitations which seem \textquotedblleft to appear from nowhere and
disappear without a trace"~\cite{Akhmediev2009are}, that makes it difficult
to systematically observe them in the ocean, where they had been originally
spotted. The origin of the RWs is still a subject of discussions \cite%
{akhmediev2010discussion}, stimulating the development of various
theoretical approaches and laboratory experiments \cite%
{yan2012rogue,henning2014taming}. In particular, it is well established that
the MI of the CW background, also known as the Benjamin-Feir instability~%
\cite{bf}, plays a pivotal role in the generation of RWs~\cite{mi1, mi2,mi3,
solli2012fluctuations}.

In addition to the basic RW solutions of the NLS equation, RWs have been
found in related integrable equations including higher-order dispersion and
nonlinearity terms, such as Hirota \cite{Ankiewicz2010rogue}, Sasa-Satsuma
\cite{Bandelow2012persistence,chen2013twisted}, Gerdjikov-Ivanov \cite%
{xu2012the,wen}, Lakshmanan-Porsezian-Daniel (LPD) \cite{wang2013breather},
quintic NLS~\cite{chowdury2015breather}, modified NLS \cite{wen}, and
Davey-Stewartson \cite{ohta2012rogue} equations, the derivative NLS
equations of the Chen-Lee-Liu and Kaup-Newell types~\cite{chan2014rogue},
coupled NLS equations~\cite{Baronio2012solutions,guo2011rogue}, three-wave
equations~\cite{threewave}, and various modifications of these models \ with
variable coefficient~\cite{yan2015two,yan2013optical,yan2015h}. Finding RWs
and rational solitons of other nonlinear wave equations in the field of
nonlinear science remains a relevant subject.

Recently, Chowdury \textit{et al.}~\cite{chowdury2014soliton} reported an
integrable three-parameter fifth-order NLS (FONLS) equation, written as
\begin{equation}
i\psi _{x}+S(\psi )-i\alpha H(\psi )+\gamma P(\psi )-i\delta Q(\psi )=0,
\label{qnls}
\end{equation}%
where $\psi \equiv \psi (x,t)$ is a complex field, subscripts denote partial
derivatives with respect to the corresponding variables, while $\alpha
,\,\gamma ,$ and $\delta $ are all real parameters. Further, $S(\psi )$
denotes the NLS part
\begin{equation}
S(\psi )=\frac{1}{2}\psi _{tt}+|\psi |^{2}\psi ,  \notag
\end{equation}%
$H(\psi )$ stands for the Hirota part,
\begin{equation}
H(\psi )=\psi _{ttt}+6|\psi |^{2}\psi _{t},  \notag
\end{equation}%
$P(\psi )$ is the LPD part,
\begin{eqnarray}
P(\psi ) &=&\psi _{tttt}+8|\psi |^{2}\psi _{tt}+6|\psi |^{4}\psi +4|\psi
_{t}|^{2}\psi  \notag \\
&&+6\psi ^{\ast }\psi _{t}^{2}+2\psi ^{2}\psi _{tt}^{\ast },  \notag
\end{eqnarray}%
and $Q(\psi )$ is the quintic part,
\begin{eqnarray}
Q(\psi ) &=&\psi _{ttttt}+10|\psi |^{2}\psi _{ttt}+10\left( \psi \left\vert
\psi _{t}\right\vert ^{2}\right) _{t}  \notag \\
&&+20\psi ^{\ast }\psi _{t}\psi _{tt}+30|\psi |^{4}\psi _{t},  \notag
\end{eqnarray}%
where $\psi ^{\ast }$ stands for the complex conjugate of $\psi $. This
equation (\ref{qnls}) is a member of the NLS hierarchy introduced by Kano~%
\cite{kano1989normal} and contains many special cases. When $\alpha =\gamma
=\delta =0$, Eq.~(\ref{qnls}) reduces to the fundamental NLS equation (\ref%
{nls}). As $\gamma =\delta =0$, it becomes the Hirota equation~\cite%
{hirota1973exact}. When $\alpha =\delta =0$, it is the LPD equation~\cite%
{lakshmanan1988effect}. All these particular nonlinear wave equations are
integrable, and the respective soliton solutions have been obtained using
different methods. Equation~(\ref{qnls}) with $\alpha =\gamma =0$ is also
integrable, and its soliton and breather solutions are available~too \cite%
{chowdury2014soliton,chowdury2015breather}.

To the best of our knowledge, rational RW solutions of Eq.~(\ref{qnls}) with
arbitrary parameters ($\alpha ,\,\gamma $, and $\delta $) have not been
reported before. In this work, we aim to produce such localized solutions
and study their dynamical properties by means of the generalized Darboux
transform.

The rest of the paper is organized as follows. In Sec. II, the Lax pair and
Darboux transform with nonzero CW background are introduced for Eq. (\ref%
{qnls}). In Sec. III, the MI of the CW states is investigated, as the
driving mechanism behind the existence of RWs. In Sec. IV, the first- and
second-order RW solutions for Eq. (\ref{qnls}) are explicitly found by means
of the Darboux transform applied to the flat CW. Trajectories of motion of
peaks and depressions of the derived RWs are explicitly considered, to
demonstrate the structure and dynamics of these patterns. Rational W-shaped
solitons of the FONLS equation, which differ from the usual bright and dark
solitons, are also found for special parameter values. The paper is
concluded by Sec. V.

\section{The Lax pair, Darboux transform, and CW solutions}

Equation (\ref{qnls}) is integrable in terms of the corresponding Lax pair,
which can be obtained as the compatibility condition, $R_{xt}=R_{tx}$,
i.e.,, $U_{x}-V_{t}+[U,V]=0$, for the following linear spectral problem~\cite%
{chowdury2014soliton}:
\begin{eqnarray}
R_{t} &=&UR,\quad U=\lambda J+U_{0},  \label{laxeq} \\
R_{x} &=&VR,\quad V=\lambda U+V_{0}+\alpha L+\gamma M+\delta N,
\label{laxeq2}
\end{eqnarray}%
where $\lambda $ is a complex spectral parameter, the Jost function $R\equiv
R(x,t)$ is a column vector given by
\begin{equation}
R=R(x,t)=\left(
\begin{array}{c}
r\vspace{0.1in} \\
s%
\end{array}%
\right) ,
\end{equation}%
with $r=r(x,t)$ and $s=s(x,t)$ being complex functions, and matrices $%
J,\,U_{0},\,V_{0},\,L,\,M$, $N$ given by
\begin{equation}
\begin{aligned} J&=\left[ \begin{array}{cc} i & 0 \\ 0 &
-i\end{array}\right] ,\vspace{0.2in}\\ U_{0}&=\left[ \begin{array}{cc} 0 &
i\psi ^{\ast } \\ i\psi & 0\end{array}\right] ,\vspace{0.2in}
\label{rjnvlms} \\ V_{0}&=\frac{1}{2}\left[ \begin{array}{cc} -i|\psi |^{2}
& \psi _{t}^{\ast }\vspace{0.1in} \\ -\psi _{t} & i|\psi
|^{2}\end{array}\right] , \\ L&=-4(\lambda ^{3}J+\lambda ^{2}U_{0}+\lambda
V_{0})+L_{0}, \\ M&=2\lambda L+M_{0}, \\ N&=-2\lambda M+N_{0}, \end{aligned}
\end{equation}%
where
\begin{equation}
\begin{aligned} L_{0}=&\left[ \begin{array}{cc} \psi \psi _{t}^{\ast }-\psi
^{\ast }\psi _{t} & i\left( 2|\psi |^{2}\psi ^{\ast }+\psi _{tt}^{\ast
}\right) \vspace{0.1in} \\ i\left( 2|\psi |^{2}\psi +\psi _{tt}\right) &
\psi ^{\ast }\psi _{t}-\psi \psi _{t}^{\ast }\end{array}\right] , \\
M_{0}=&\left[ \begin{array}{cc} m_{1} & m_{2} \\ m_{3} &
-m_{1}\end{array}\right] ,\vspace{0.1in} \\ N_{0}=&\left[ \begin{array}{cc}
n_{1} & n_{2} \\ n_{3} & -n_{1}\end{array}\right] ,\\ m_{1}=&-i\left( 3|\psi
|^{4}+\psi \psi _{tt}^{\ast }-\psi _{t}\psi _{t}^{\ast }+\psi ^{\ast }\psi
_{tt}\right) , \label{rjnvlms2} \\ m_{2}=&-m_{3}^{\ast } \\ =& 6|\psi
|^{2}\psi _{t}^{\ast}+\psi _{ttt}^{\ast }, \\ n_{1}=&(\psi -\psi ^{\ast
})\psi _{ttt}-\psi _{t}\psi _{tt}^{\ast }+\psi _{t}^{\ast }\psi _{tt} \\
&+6|\psi |^{2}(\psi \psi _{t}^{\ast }-\psi ^{\ast }\psi _{t}), \\
n_{2}=&-n_{3}^{\ast } \\ =& 6i|\psi|^{4}\psi ^{\ast }+2i\psi ^{\ast 2}\psi
_{tt}+4i|\psi _{t}|^{2}\psi ^{\ast }+6i\psi \psi _{t}^{\ast 2} \\ & +8i|\psi
|^{2}\psi _{tt}^{\ast }+i\psi _{tttt}^{\ast }. \end{aligned}
\end{equation}

Based on the Lax pair of an integrable nonlinear wave equation, the Darboux
transform is a powerful method for generating soliton and breather
solutions, using a zero solution or a flat CW as the \textquotedblleft seed"
solution~\cite{dt}. The hierarchy of multisoliton states~\cite%
{akhmediev1991extremely} can be derived starting from zero, while solutions
related to the MI~\cite{akhmediev2009extreme} are obtained starting from the
CW.

The Darboux transform for Eq.~(\ref{qnls}) is given by
\begin{equation}
\psi _{n+1}(x,t)=\psi _{n}(x,t)-\frac{4is_{n+1}(x,t)r_{n+1}^{\ast }(x,t)}{%
|r_{n+1}(x,t)|^{2}+|s_{n+1}(x,t)|^{2}},\,\,\,  \label{gdt}
\end{equation}%
where $n=0,1,2,...$, $\psi _{n}(x,t)$ is a seed solution of Eq.~(\ref{qnls}%
), while $\psi _{n+1}(x,t)$ is a new solution generated from $\psi _{n}(x,t)$%
, while $\ r_{n+1}(x,t)$ and $s_{n+1}(x,t)$ are solutions of the linear
spectral problem based on Eqs.~(\ref{laxeq})-(\ref{rjnvlms2}), with $\psi
(x,t)$ substituted by $\psi _{n+1}(x,t)$. We here choose $\lambda =i$ in
system ~(\ref{laxeq})-(\ref{rjnvlms2}), as this value of the spectral
parameter corresponds to RW solutions of Eq.~(\ref{qnls}).

The zero seed solution was employed to build multisoliton solutions of Eq.~(%
\ref{qnls}) in Ref.~\cite{chowdury2014soliton}. In order to use the Darboux
transform (\ref{gdt}) for obtaining RW solutions of Eq.~(\ref{qnls}), we
here take the CW solution of Eq.~(\ref{qnls}),
\begin{equation}
\psi _{0}(x,t)=ae^{i(bx+ct)},  \label{planwav}
\end{equation}%
as the seed, where $a\not=0$ and $c$ are arbitrary real constants, and
\begin{equation}
\begin{aligned} b=\alpha c\left( 6a^{2}-c^{2}\right) +\gamma \left(
6a^{4}-12a^{2}c^{2}+c^{4}\right) \\ +c\delta \left(
30a^{4}-20a^{2}c^{2}+c^{4}\right) -\frac{c^{2}}{2}+a^{2}. \label{rbc}
\end{aligned}
\end{equation}

Lastly, we note that FONLS equation (\ref{qnls}) is invariant under the
following scale transformation:
\begin{equation}
\begin{aligned} x\rightarrow \mu ^{2}x,\quad t\rightarrow \mu t, \quad \psi
\rightarrow \frac{\psi }{\mu },\\ \quad \alpha \rightarrow \mu \alpha ,\quad
\gamma \rightarrow \mu ^{2}\gamma ,\quad \delta \rightarrow \mu ^{3}\delta ,
\end{aligned}  \notag
\end{equation}%
where $\mu $ is an arbitrary real constant. We stress that this invariance
is different from that for the NLS equation~\cite{nls}.

\section{The modulational instability (MI) of CW states}

To investigate the MI of the CW solution (\ref{planwav}), subject to
constraint (\ref{rbc}), we add perturbations $\phi \left( x,t\right) $ with
an infinitesimal amplitude $\varepsilon $,
\begin{equation}
\hat{\psi}(x,t)=[a+\varepsilon \phi (x,t)]e^{i(bx+ct)},
\end{equation}%
which is followed by the derivation of the linearized equation for the
perturbation:
\begin{equation}
\begin{array}{l}
i\phi _{x}-i\delta \phi _{ttttt}+(\gamma +5c\delta )\phi
_{tttt}+i(10c^{2}-10a^{2}\delta \vspace{0.1in} \\
\quad +4c\gamma -\alpha )\phi _{ttt}+(40a^{2}c\delta +3c\alpha -6c^{2}\gamma
-10c^{3}\delta \vspace{0.1in} \\
\quad +8a^{2}\gamma +1/2)\phi _{tt}+2a^{2}(5c\delta +\gamma )\phi
_{tt}^{\ast }+i(-30a^{4}\qquad \vspace{0.1in} \\
\quad +60a^{2}c^{2}\delta -5c^{4}\delta +24a^{2}c\gamma -4c^{3}\gamma
-6a^{2}\alpha \vspace{0.1in} \\
\quad +3c^{2}\alpha +c)\phi _{t}+a^{2}(60a^{2}c\delta -20c^{3}\delta
+12a^{2}\gamma \vspace{0.1in} \\
\quad -12c^{2}\gamma +6c\alpha +1)(\phi +\phi ^{\ast })=0.%
\end{array}
\label{linear}
\end{equation}

Relevant solutions to Eq. (\ref{linear}) with wavenumber $k$ and frequency $%
\omega $ are sought for as
\begin{equation}
\phi (x,t)=F\cos (kx-\omega t)+iG\sin (kx-\omega t),  \label{phisol}
\end{equation}%
where $F$ and $\,G$ are real amplitudes. The substitution of ansatz (\ref%
{phisol}) into Eq. (\ref{linear}) readily leads to the following dispersion
relation for the perturbations, obtained as the condition for the existence
of nontrivial solutions for $F$ and $G$:
\begin{equation}
\begin{array}{l}
4\left\{ k-\omega \left[ c+\alpha (3c^{2}-6a^{2}+k^{2})+4\gamma
c(6a^{2}-c^{2}-k^{2})\right. \right. \quad \vspace{0.1in} \\
\left. \left. +5\delta (12a^{2}c^{2}-6a^{4}+2a^{2}k^{2}-2c^{2}\omega
^{2}-c^{4}-\omega ^{4})\right] \right\} ^{2}\vspace{0.1in} \\
\qquad =-\omega ^{2}(4a^{2}-\omega ^{2})\left[ 10c\delta
(6a^{2}-2c^{2}-\omega ^{2})\right. \vspace{0.1in} \\
\qquad \qquad \left. +2\gamma (6a^{2}-6c^{2}-\omega ^{2})+6\alpha c+1\right]
^{2}.%
\end{array}
\label{relation}
\end{equation}%
It followsf rom Eq.~(\ref{relation}) that we know the wavenumber $k$ is
complex for frequencies interval
\begin{equation}
|\omega |<2|a|,  \notag
\end{equation}%
hence the CW states are subject to the MI, and RW patterns may be expected
as solutions of Eq.~(\ref{qnls}) in this case.

\section{Rogue waves, rational solitons, and the dynamical analysis}

\subsection{First-order rational solutions and trajectories of extreme points%
}

We choose $a=1$ and $c=0$ in the seed solution given by Eqs.~(\ref{planwav})
and~(\ref{rbc}), i.e.,
\begin{equation}
\psi _{0}(x,t)=e^{i(6\gamma +1)x},  \label{seed}
\end{equation}%
Solving the linear spectral problem ~(\ref{laxeq})-(\ref{rjnvlms2}) with $%
\lambda =i$ and the seed solution (\ref{seed}) yields eigenfunctions
\begin{equation}
\begin{array}{l}
r_{1}(x,t)=-c_{0}[2(t+vx+iBx)-1]e^{-i(B+1)x/4},\vspace{0.1in} \\
s_{1}(x,t)=c_{0}[2(t+vx+iBx)+1]e^{i(B+1)x/4},\qquad%
\end{array}
\label{rs1}
\end{equation}%
where $c_{0}$ is an arbitrary nonzero constant, and $v,B$ are given by
\begin{equation}
B=12\gamma +1,\qquad v=6(\alpha +5\delta ).  \label{rdbv}
\end{equation}

Then, substituting solution (\ref{seed}) and eigenfunctions (\ref{rs1}) into
the Darboux transform (\ref{gdt}), we obtain the first-order rational
solution of Eq.~(\ref{qnls}) as
\begin{equation}
\psi _{1}(x,t)=\left[ 1-\frac{4(1+2iBx)}{D_{1}(x,t)}\right] \exp \left(
\frac{i}{2}(B+1)x\right) ,  \label{rog1sol}
\end{equation}%
where we define
\begin{equation}
D_{1}(x,t)=4(t+vx)^{2}+4B^{2}x^{2}+1,  \label{rdbv2}
\end{equation}

From the first-order rational solution (\ref{rog1sol}) of Eq.~(\ref{qnls}),
the corresponding first-order RW solutions can be obtained also for the
Hirota, LPD, and quintic NLS equations, by choosing the respective values of
$\alpha ,\,\gamma $, and $\delta $, as specified in Introduction. For
example, for (i) $\alpha =\delta =\gamma =0$, we also find the known
first-order RW solution of the self-focusing NLS equation,
\begin{equation}
\psi _{\mathrm{1,NLS}}=\left[ 1-\frac{4(1+2ix)}{4(t^{2}+x^{2})+1}\right]
e^{ix};  \notag
\end{equation}%
(ii) for $\delta =\gamma =0$, we obtain the first-order RW solution of the
Hirota equation,
\begin{equation}
\psi _{\mathrm{1,Hirota}}=\left[ 1-\frac{4(1+2ix)}{4(t+6\alpha
x)^{2}+4x^{2}+1}\right] e^{ix};  \notag
\end{equation}%
(iii) for $\alpha =\delta =0$, the first-order RW solution of the LPD
equation is obtained:
\begin{equation}
\psi _{\mathrm{1,LPD}}\!=\!\left\{ 1-\frac{4[1+2i(12\gamma +1)x]}{%
4t^{2}+4(12\gamma +1)^{2}x^{2}+1}\right\} \!e^{i(6\gamma +1)x}.\,\,  \notag
\end{equation}

In the following we will analyze the solution (\ref{rog1sol}) for two
different cases $\gamma\not=-1/12$ and $\gamma=-1/12$, which display the
distinguish profiles of solution (\ref{rog1sol}).

\subsubsection{The first-order RW solution with $\protect\gamma \not=-1/12$}

With $\gamma \not=-1/12$, the rational solution (\ref{rog1sol}) of Eq.~(\ref%
{qnls}) is just its the first-order RW solution. To analyze dynamical
properties of the RW solution, we consider its intensity,
\begin{equation}
|\psi _{1}(x,t)|^{2}=\frac{8-32(t+vx)^{2}+32B^{2}x^{2}}{%
[1+4(t+vx)^{2}+4B^{2}x^{2}]^{2}}+1.  \label{amp}
\end{equation}%
It is easy to see that expression (\ref{amp}) has three critical points,
\begin{equation}
(x_{1},t_{1})=(0,0),\quad (x_{2,3},t_{2,3})=\left( 0,\pm \frac{\sqrt{3}}{2}%
\right) ,  \notag
\end{equation}%
which are three real roots of system $\{\partial \left( |\psi
_{1}(x,t)|^{2}\right) /\partial t=0,\,\partial \left( |\psi
_{1}(x,t)|^{2}\right) /\partial x=0\}$ (see Appendix A). Further, the
critical point $(x_{1},t_{1})$ is a maximum (peak), as
\begin{equation}
\frac{\partial ^{2}\left( |\psi _{1}(x,t)|^{2}\right) }{\partial t^{2}}%
=-192<0,  \notag
\end{equation}%
and
\begin{eqnarray}
&&\left( \frac{\partial ^{2}\left( |\psi _{1}(x,t)|^{2}\right) }{\partial
x\partial t}\right) ^{2}-\frac{\partial ^{2}\left( |\psi
_{1}(x,t)|^{2}\right) }{\partial t^{2}}\frac{\partial ^{2}\left( |\psi
_{1}(x,t)|^{2}\right) }{\partial x^{2}}  \notag \\
&=&-12288(12\gamma +1)^{2}<0  \label{xt1}
\end{eqnarray}%
at point $(x_{1},t_{1})$, which makes the meaning of condition $\gamma
\not=-1/12$ obvious. Similarly, two critical points $(x_{2,3},t_{2,3})$ are
minima (depressions), as
\begin{equation}
\frac{\partial ^{2}\left( |\psi _{1}(x,t)|^{2}\right) }{\partial t^{2}}=6>0,
\notag
\end{equation}%
and
\begin{eqnarray}
&&\left( \frac{\partial ^{2}\left( |\psi _{1}(x,t)|^{2}\right) }{\partial
x\partial t}\right) ^{2}-\frac{\partial ^{2}\left( |\psi
_{1}(x,t)|^{2}\right) }{\partial t^{2}}\frac{\partial ^{2}\left( |\psi
_{1}(x,t)|^{2}\right) }{\partial x^{2}}  \notag \\
&=&-48(12\gamma +1)^{2}<0  \label{xt2}
\end{eqnarray}%
at points $(x_{2,3},t_{2,3})$.
\begin{figure}[!t]
\centering
\vspace{0.1in}  {\scalebox{0.38}[0.38]{\includegraphics{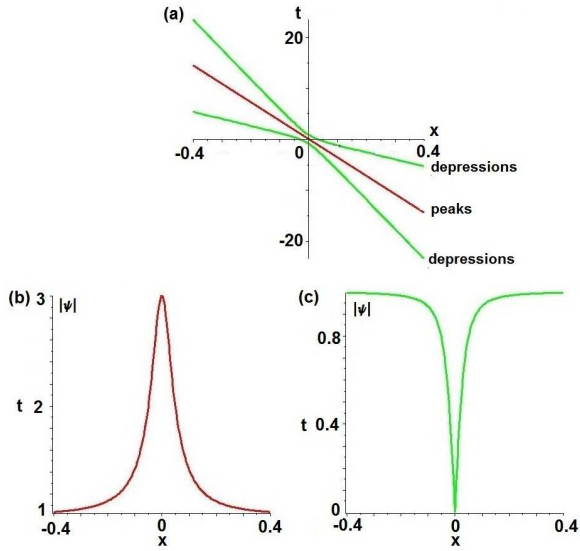}}}
\caption{ {\protect\small (Color online). The first-order rogue wave (%
\protect\ref{rog1sol}) for $\protect\alpha =\protect\gamma=\protect\delta =1$%
: (a) The motion of the peak and depression centers. (b) The evolution of
the peak absolute value. (c) The evolution of the absolute value of the
field at the depression points.}}
\label{rog1traj-fig}
\end{figure}

\begin{figure}[!t]
\centering
\vspace{0.1in} {\scalebox{0.38}[0.38]{\includegraphics{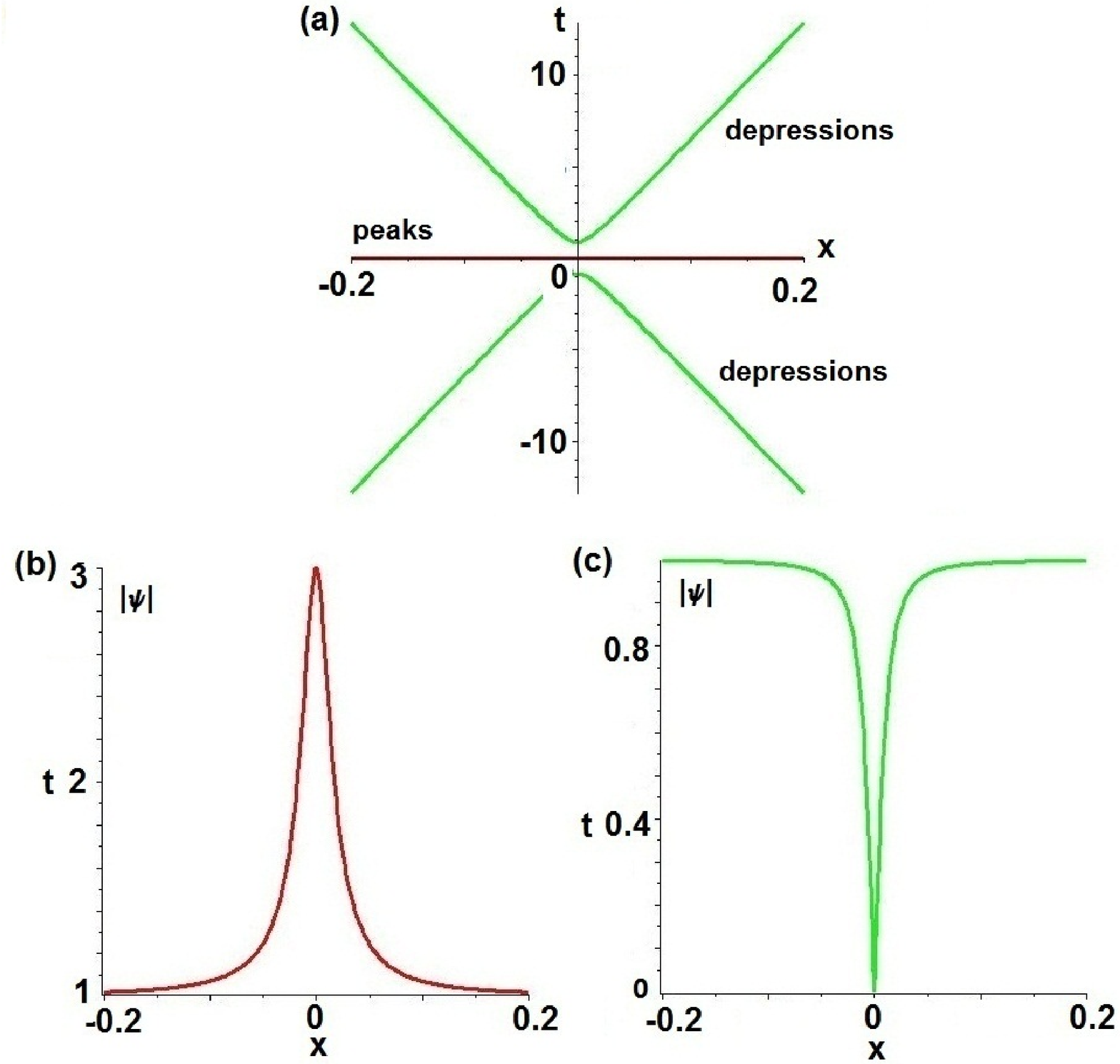}}}
\caption{ {\protect\small (Color online). The same as in Fig. \protect\ref%
{rog1traj-fig}, but for parameters $\protect\alpha =-5,\,\protect\gamma =3$,
and $\protect\delta =1$.}}
\label{rog1(v=0)traj-fig}
\end{figure}

\begin{figure}[!t]
\centering
\vspace{0.1in} \includegraphics[width=3.2in]{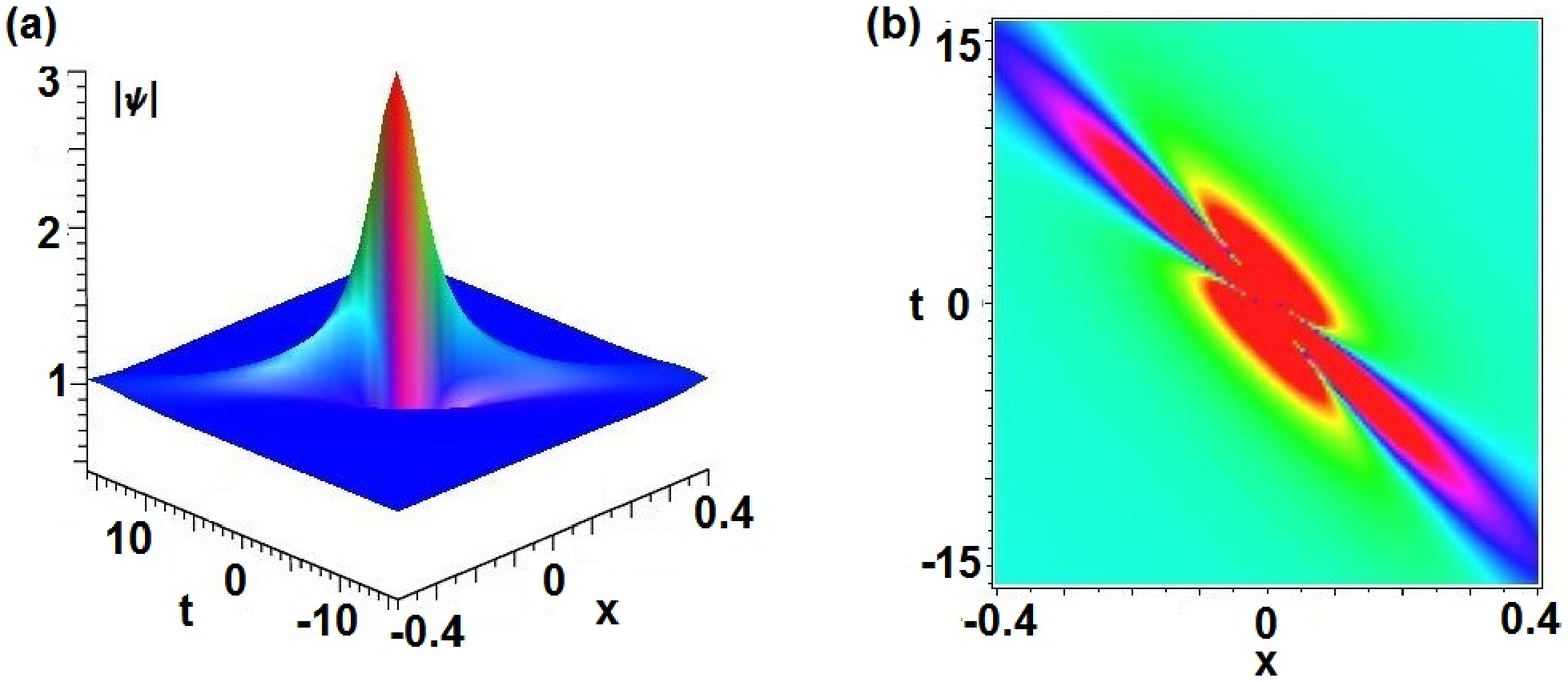}
\caption{ {\protect\small (Color online). (a) The density evolution and (b)
2D density contour plots for the first-order RW solution (\protect\ref%
{rog1sol}) at $\protect\alpha =\protect\gamma =\protect\delta =1$. The
corresponding motion of peaks and depressions is displayed in Fig.~\protect
\ref{rog1traj-fig}.}}
\label{rog1-fig}
\end{figure}

\begin{figure}[!t]
\centering
\vspace{0.1in} \includegraphics[width=3.2in]{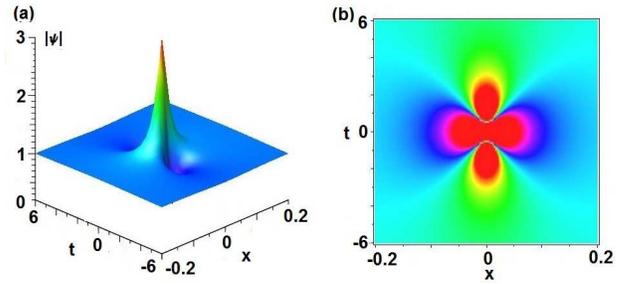}
\caption{ {\protect\small (Color online). (a) The density evolution and (b)
2D density contour plots for the first-order RW solution (\protect\ref%
{rog1sol}) at $\protect\alpha =-5,\,\protect\gamma =3$ and $\protect\delta %
=1 $. The corresponding motion of peaks and depressions is displayed in Fig.~%
\protect\ref{rog1(v=0)traj-fig}.}}
\label{rog1(v=0)-fig}
\end{figure}
Note that these critical points, $(x_{j},t_{j})$,~$\,j=1,2,3$, do not depend
on parameters $\alpha ,\,\gamma $, and $\delta $.

The evolution of the first-order RW solution can be characterized by
trajectories of peaks and depressions~\cite{ling2013simple}. The trajectory
of the peak's center is given by the dependence of its temporal coordinate, $%
T_{h}$, on $x$. It follows from Eq. (\ref{rog1sol}) that
\begin{equation}
T_{h}=-vx,  \label{rog1hump}
\end{equation}%
and the temporal coordinates, $T_{c\pm }$, of centers of the two depressions
are given by
\begin{equation}
T_{c\pm }=-vx\pm \frac{1}{2}\sqrt{3(1+4B^{2}x^{2})}.  \label{rog1valley}
\end{equation}%
Thus, the motion of the peak is totally determined by parameter $v=6(\alpha
+5\delta )$ defined in Eq. (\ref{rdbv}), while the RW width, which is
defined as the temporal distance between the two depressions, i.e.,
\begin{equation}
T_{d}=\sqrt{3(1+4B^{2}x^{2})},  \notag
\end{equation}%
is determined by parameter $B=12\gamma +1$, which is also defined by Eq. (%
\ref{rdbv}). Thus, from the full set of three constants $\alpha ,\gamma ,$
and $\delta $ in Eq.~(\ref{qnls}), two parameters, $\alpha $ and $\delta $,
control the motion of the RW peak in solution (\ref{rog1sol}), while the
third parameter, $\gamma $, determines the temporal distance between the two
depressions.

To illustrate the structure and dynamical properties of the first-order RW
solution (\ref{rog1sol}) at different parameters, we choose sets $\left(
\alpha =1,\,\gamma =1,\,\delta =1\right) $ and $\left( \alpha =5,\,\gamma
=3,\,\delta =1\right) $ to display traces of peak (\ref{rog1hump}) and
depressions (\ref{rog1valley}) in Figs. \ref{rog1traj-fig}(a) and \ref%
{rog1(v=0)traj-fig}(a). Further, the evolution of the absolute value of the
field at the peak and depression points is shown in Figs. \ref{rog1traj-fig}%
(b,c) and \ref{rog1(v=0)traj-fig}(b,c).

Lastly, 2D density contour plots of the RW solution (\ref{rog1sol}) are
presented in Figs.~\ref{rog1-fig} and \ref{rog1(v=0)-fig} for different
parameters.

\begin{figure}[!t]
\centering
\vspace{0.1in} \includegraphics[width=2.75in]{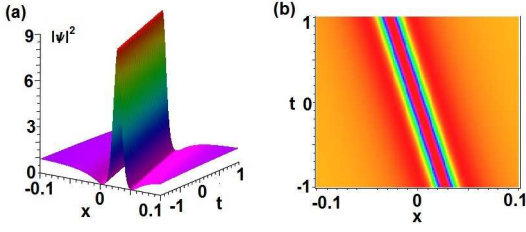}
\caption{(a) The evolution of the W-shaped profile and (b) 2D density
contour plot for rational soliton (\protect\ref{soliton}) with $\protect%
\alpha =1,\,\protect\gamma =-1/12$ and $\protect\delta =1$.}
\label{fig-soliton}
\end{figure}

\begin{figure}[!t]
\centering
\vspace{0.1in} \includegraphics[width=2.75in]{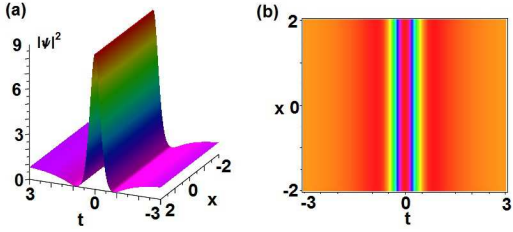}
\caption{(a) The evolution of the W-shaped intensity profile and (b) 2D
intensity contour plot for rational soliton (\protect\ref{soliton}) at $%
\protect\alpha =-5,\,\protect\gamma =-1/12$ and $\protect\delta =1$.}
\label{fig2-soliton}
\end{figure}

\subsubsection{The first-order rational soliton with $\protect\gamma =-1/12$}

For the case of $\gamma =-1/12$, we find that rational solution (\ref%
{rog1sol}) becomes a solitary-wave solution of Eq.~(\ref{qnls}),
\begin{equation}
\psi _{1\gamma }(x,t)=\left[ 1-\frac{4}{1+4(t+vx)^{2}}\right] e^{ix/2},
\label{soliton}
\end{equation}%
which is a rational soliton-like solution of (\ref{qnls}) with properties
\begin{equation}
|\psi _{1\gamma }(x,t)|^{2}\rightarrow 1\quad \mathrm{as}\quad
|t+vx|\rightarrow \infty ,  \notag
\end{equation}%
\begin{equation}
|\psi _{1\gamma }(x,t)|^{2}\rightarrow 9\quad \mathrm{as}\quad
t+vx\rightarrow 0,  \notag
\end{equation}%
\begin{equation}
\mathrm{max}(|\psi _{1\gamma }(x,t)|^{2})=9\quad \mathrm{as}\quad t+vx=0,
\notag
\end{equation}%
\begin{equation}
\mathrm{min}(|\psi _{1\gamma }(x,t)|^{2})=0\quad \mathrm{as}\quad |t+vx|=%
\sqrt{3}/2,  \notag
\end{equation}%
Rational soliton (\ref{soliton}) feature W-shaped profiles at all $x$, as
shown in Figs.~\ref{fig-soliton} and ~\ref{fig2-soliton}.

Next, we analyze critical points of the intensity of the rational solution (%
\ref{soliton}). It follows from Eq. (\ref{soliton}) that there are three
critical points of $\left\vert \psi _{1\gamma }\right\vert ^{2}$,
\begin{equation}
(\tilde{x}_{1},\tilde{t}_{1})\!=\!(x,\,-vx),\,\,(\tilde{x}_{2,3},\tilde{t}%
_{2,3})\!=\!\left( x,\,-vx\pm \frac{\sqrt{3}}{2}\right) ,\quad
\label{s-ponit}
\end{equation}%
whose trajectories are straight lines. It also follows from Eq.~(\ref%
{soliton}) that
\begin{equation}
\left(\!\! \frac{\partial ^{2}\left( |\psi_{1\gamma }(x,t)|^{2}\right) }{%
\partial x\partial t}\!\!\right) ^{2}\!-\!\frac{\partial ^{2}\left( |\psi
_{1\gamma }(x,t)|^{2}\right) }{\partial t^{2}}\frac{\partial ^{2}\left(
|\psi _{1\gamma }(x,t)|^{2}\right) }{\partial x^{2}}\!=\! 0,  \notag
\end{equation}%
at three critical points (\ref{s-ponit}) [in agreement with Eqs. (\ref{xt1})
and (\ref{xt2})], hence the points cannot be characterized on the basis of
this equation.

On the other hand, the critical point $(\tilde{x}_{1},\tilde{t}_{1})$ is a
maximum (peak), as
\begin{equation}
\frac{\partial ^{2}\left( |\psi _{1\gamma }(x,t)|^{2}\right) }{\partial t^{2}%
}=-192<0  \notag
\end{equation}%
at this point (see Figs.~\ref{fig-soliton} and \ref{fig2-soliton}).
Similarly, two critical points $(\tilde{x}_{2,3},\tilde{t}_{2,3})$ are
minima (depressions), as
\begin{equation}
\frac{\partial ^{2}\left( |\psi _{1\gamma }(x,t)|^{2}\right) }{\partial t^{2}%
}=6>0  \notag
\end{equation}%
at these points (see Figs.~\ref{fig-soliton} and \ref{fig2-soliton}).

\subsection{Second-order rational solutions and trajectories of extreme
points}

To produce second-order RW solution of Eq.~(\ref{qnls}), we choose the
first-order rational solution (\ref{rog1sol}) found above as a new seed.
Substituting it into the linear spectral problem based on Eqs.~(\ref{laxeq}%
)-(\ref{rjnvlms2}) with $\lambda =i$, we solve it to generate new complex
functions $r_{2}(x,t)$ and $s_{2}(x,t)$,
\begin{eqnarray}  \label{rrs}
\begin{array}{rcl}
r_{2}(x,t) & = & \left[ r_{21}(x,t)+ir_{22}(x,t)\right] e^{-i(B+1)x/4},
\vspace{0.1in} \\
s_{2}(x,t) & = & \left[ s_{21}(x,t)+is_{22}(x,t)\right] e^{i(B+1)x/4},%
\end{array}%
\end{eqnarray}
where $r_{2j}(x,t)$ and $s_{2j}(x,t)$ \, $(j=1,2)$ are given by
\begin{equation}
\begin{array}{rl}
r_{21}(x,t) & =\dfrac{c_{1}\xi (x,t)+c_{2}B\eta (x,t)}{4D_{1}(x,t)},\vspace{%
0.1in} \\
r_{22}(x,t) & =\dfrac{-4c_{2}B\xi (x,t)+c_{1}\eta (x,t)}{8D_{1}(x,t)},%
\vspace{0.1in} \\
s_{21}(x,t) & =\dfrac{4c_{2}B\xi (-x,-t)-c_{1}\eta (-x,-t)}{8D_{1}(x,t)},%
\vspace{0.1in} \\
s_{22}(x,t) & =\dfrac{c_{1}\xi (-x,-t)+c_{2}B\eta (-x,-t)}{4D_{1}(x,t)},%
\end{array}
\label{rrssol2}
\end{equation}
\begin{equation}
\begin{array}{rl}
\xi (x,t)\equiv & 16Bv(B^{2}+v^{2})x^{4}+16B(B^{2}+3v^{2})x^{3}t\vspace{0.1in%
} \\
& -16B^{3}x^{3}+48Bvx^{2}t^{2}+16Bxt^{3}\vspace{0.1in} \\
& +4(120B\delta -3Bv+4v)x^{2}\vspace{0.1in} \\
& +(8-20B)x+4(4-7B)xt,\vspace{0.1in} \\
\eta (x,t)\equiv & 16t^{3}(1-t-4vx)+48(B^{2}+v^{2})x^{2}t\vspace{0.1in} \\
& +12t-96v^{2}x^{2}t^{2}+32(30\delta +v)xt\vspace{0.1in} \\
& +4(7v+120\delta )x+16(B^{4}-v^{4})x^{4}\vspace{0.1in} \\
& +16v(3B^{2}+v^{2})x^{3}\vspace{0.1in} \\
& +8(7B^{2}+120\delta +4v^{2}-4B),%
\end{array}
\label{rxieta2}
\end{equation}%
with expressions $B,\,v,\,D_{1}(x,t)$ given by Eqs.~(\ref{rdbv}) and (\ref%
{rdbv2}).

\begin{figure}[!t]
\centering
\includegraphics[width=3.4in]{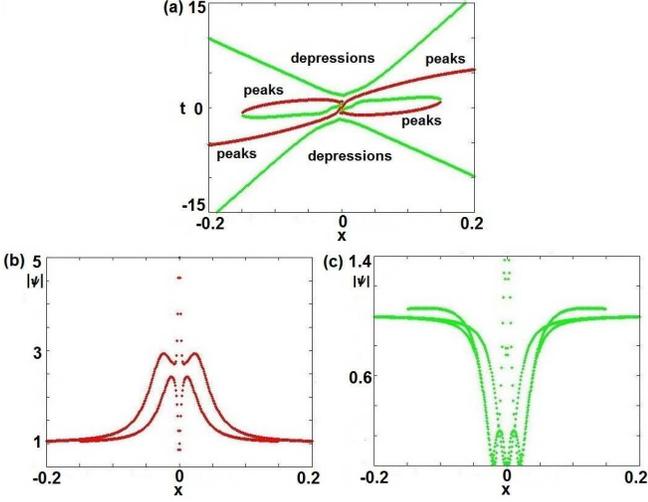}
\caption{ {\protect\small (color online). The second-order rogue wave
solution ( \protect\ref{rog2sol}) for $\protect\alpha =2,\,\protect\gamma =3$
and $\protect\delta =-1$. (a): The motion of the peak and depressions (red
and green lines, respectively). (b) and (c): The evolution of the absolute
values corresponding, severally, to the peak and depressions.}}
\label{rog2traj-fig}
\end{figure}

\begin{figure}[!t]
\centering
\vspace{0.1in}\includegraphics[width=3.4in]{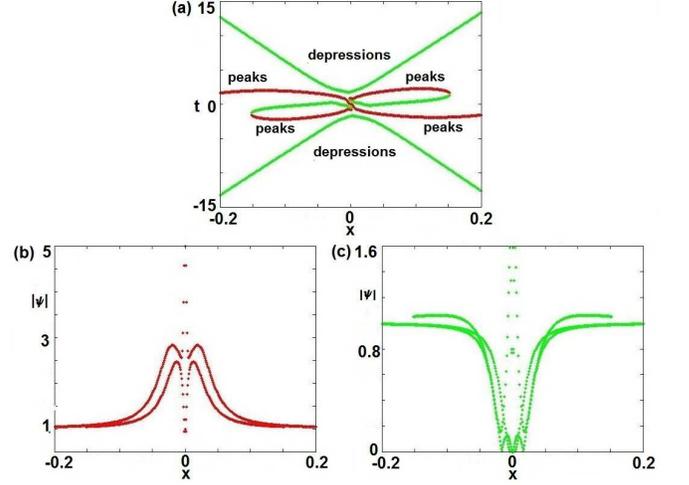}
\caption{ {\protect\small (color online). The same as in Fig.~ \protect\ref%
{rog2traj-fig}, but for parameters $\protect\alpha =-5,\, \protect\gamma =3$%
, and $\protect\delta =1$.}}
\label{rog2(v=0)traj-fig}
\end{figure}

Substituting expressions~(\ref{rrssol2}) and (\ref{rxieta2}) into system (%
\ref{rrs}) and using Darboux transform (\ref{gdt}) yields exact second-order
rational solutions of Eq.~(\ref{qnls}) as
\begin{equation}
\psi _{2}(x,t)=\left[ 1+\frac{G_{2}(x,t)+iK_{2}(x,t)}{D_{2}(x,t)}\right]
\exp \left( \frac{i}{2}(1+B)x\right) .  \label{rog2sol}
\end{equation}%
Cumbersome expressions for real functions $D_{2}(x,t)$,\thinspace\ $%
G_{2}(x,t)$, and $K_{2}(x,t)$ are given in the Appendix B. Particularly, for
cases (i) $\alpha =\delta =\gamma =0$, (ii) $\delta =\gamma =0$, and (iii) $%
\alpha =\gamma =0$, solution (\ref{rog2sol}) yields the second-order RW
solutions of NLS, Hirota, and LPD equations, respectively.

\subsubsection{The second-order RW solution with $\protect\gamma \not=-1/12$}

To analyze dynamical properties of the second-order RW solution (\ref%
{rog2sol}) with $\gamma \not=-1/12$, we first consider its intensity in the
form of
\begin{equation}
|\psi _{2}(x,t)|^{2}=\frac{[D_{2}(x,t)+G_{2}(x,t)]^{2}+K_{2}^{2}(x,t)}{%
D_{2}^{2}(x,t)}.  \label{amp2}
\end{equation}

\begin{figure}[!t]
\centering
\vspace{0.1in} \includegraphics[width=3.15in]{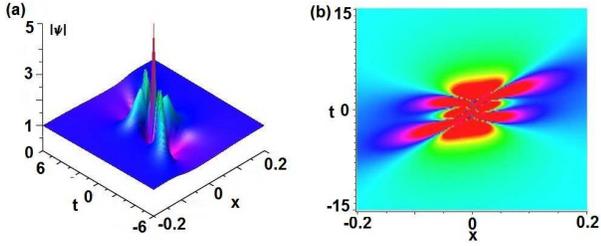}
\caption{ {\protect\small (color online). The second-order rogue wave
solution ( \protect\ref{rog2sol}) at $\protect\alpha =2,\,\protect\gamma =3$%
, and $\protect\delta =-1$. (a): The density evolution; (c) the
corresponding 2D contour plot.}}
\label{rog2-fig}
\end{figure}

\begin{figure}[!t]
\centering
\vspace{0.1in} \includegraphics[width=3.15in]{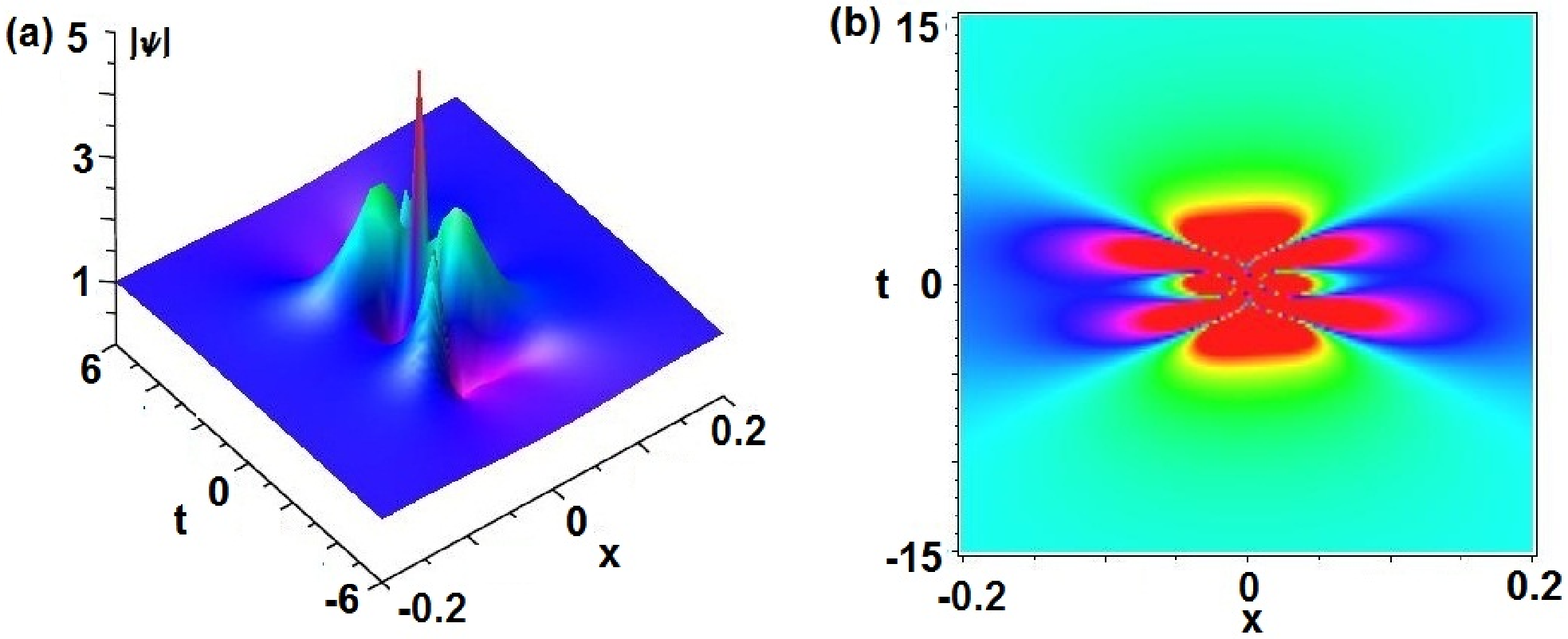}
\caption{ {\protect\small (color online). The second-order rogue wave
solution ( \protect\ref{rog2sol}) at $\protect\alpha =-5,\,\protect\gamma =3$%
, and $\protect\delta =1$. (a): The density evolution; (c) the corresponding
2D contour plot.}}
\label{rog2(v=0)-fig}
\end{figure}

Unlike the first-order rational solution (\ref{rog1sol}), the peak and
depressions of $|\psi _{2}(x,t)|^{2}$ cannot be obtained in an analytical
form for expression (\ref{amp2}) since we need to analytically solve two
algebraic equations of the eleventh order for $x$ and $t$, which are
generated by $\{\partial (|\psi _{2}(x,t)|^{2})/\partial x=0,\,\partial
|\left( \psi _{2}(x,t))|^{2}\right) /\partial t=0\}$. Therefore, similar to
what was done above for the first-order rational solution, we numerically
analyze the respective trajectories with $\gamma \not=-1/12$ (e.g., $\gamma
=3$). The results are presented in Figs.~\ref{rog2traj-fig} and \ref%
{rog2(v=0)traj-fig}, and the corresponding 2D contour plots are displayed in
Figs. \ref{rog2-fig} and \ref{rog2(v=0)-fig}.

A characteristic difference of the second-order RW pattern from its
first-order counterpart is that, in a limited interval of the evolution
variable [namely, $|x|<0.149$ and $|x|<0.152$ in Figs. \ref{rog2traj-fig}(a)
and \ref{rog2(v=0)traj-fig}(a), respectively], in addition to the set of the
single peak and two depressions, a new peak-depression pair appears.

\begin{figure}[!t]
\centering
\vspace{0.1in} \includegraphics[width=3.4in]{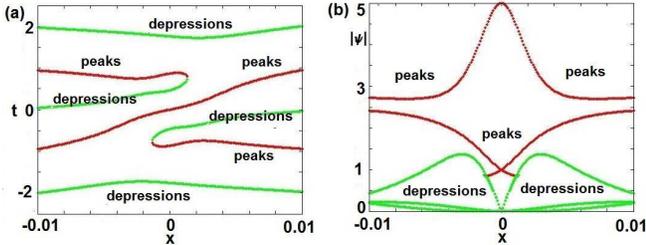}
\caption{ {\protect\small (Color online). The second-order rogue wave
solution (\protect\ref{rog2sol}) for $\protect\alpha =2,\,\protect\gamma =3$%
, and $\protect\delta =-1$ in the vicinity of $x=0$. (a): The motion of
peaks and depressions (red and green lines, respectively). (b) The evolution
of the absolute values of the peaks and depressions (red and green lines,
respectively).}}
\label{vicinity-x=0}
\end{figure}

\begin{figure}[!t]
\centering
\vspace{0.1in} \includegraphics[width=3.2in]{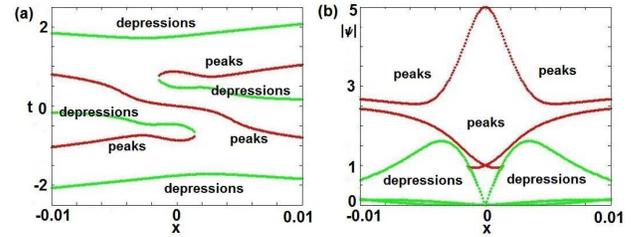}
\caption{ {\protect\small (Color online). The second-order rogue wave
solution (\protect\ref{rog2sol}) for $\protect\alpha =-5,\,\protect\gamma =3$%
, and $\protect\delta =1$. (a): The motion of the peaks and depressions (red
and green lines, respectively); (b) The evolution of the absolute values of
the peaks and depressions (red and green lines, respectively).}}
\label{vicinity-x=0(v=0)}
\end{figure}

To better represent the relevant solutions close to $x=0$, trajectories of
the peaks and depressions, and the evolution of the respective absolute
values in the interval of $|x|\leq 0.01$ are displayed, for different
parameters, in Figs. \ref{vicinity-x=0} and \ref{vicinity-x=0(v=0)}. In the
case of $\alpha =2,\,\gamma =3$ and $\delta =-1$, there are two peaks and
three depressions in the intervals of $-0.149\leq x\leq \,-0.001$ and $%
0.001\leq x\leq 0.149$, and there are three peaks and four depressions in
the interval of $|x|\leq 0.001$. In the case of $\alpha =-5,\,\gamma =3$ and
$\delta =1$, there are two peaks and three depressions in the intervals of $%
-0.152\leq x\leq -0.001$ and $0.001\leq x\leq 0.152$, and there are three
peaks and four depressions at $|x|\leq 0.001$.


\subsubsection{Rational soliton pair with $\protect\gamma =-1/12$}

For $\gamma =-1/12$, Eq. (\ref{rog2sol}) yields rational two-soliton
solutions of Eq.~(\ref{qnls}) as
\begin{equation}
\psi _{2\gamma }(x,t)=\left[ 1+\frac{G_{2\gamma }(x,t)+iK_{2\gamma }(x,t)}{%
D_{2\gamma }(x,t)}\right] e^{ix/2}.  \label{soliton2}
\end{equation}%
Cumbersome expressions for real functions $D_{2\gamma }(x,t)$,\thinspace\ $%
G_{2\gamma }(x,t)$, and $K_{2\gamma }(x,t)$ are given in the Appendix C.

\begin{figure}[!ht]
\centering
\vspace{0.1in} \includegraphics[width=3in]{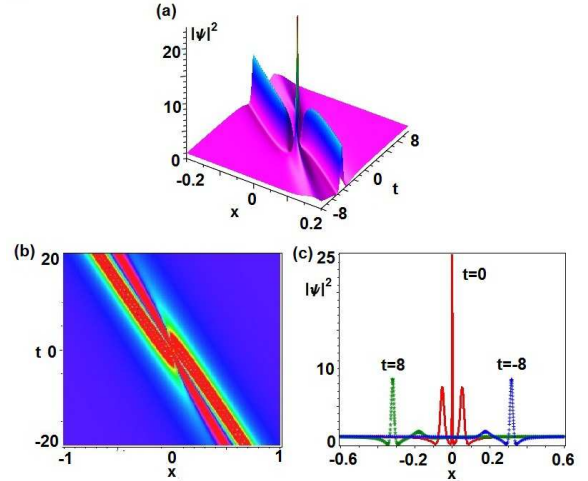}
\caption{(a) The interaction of two W-shaped solitons (\protect\ref{soliton2}%
); (b) 2D density contour plots; and (c) cross sections for two-solitons (%
\protect\ref{soliton2}), for $\protect\alpha =\protect\delta=1$, and $%
\protect\gamma =-1/12$.}
\label{fig-2soliton}
\end{figure}

\begin{figure}[!ht]
\centering
\vspace{0.1in} \includegraphics[width=3in]{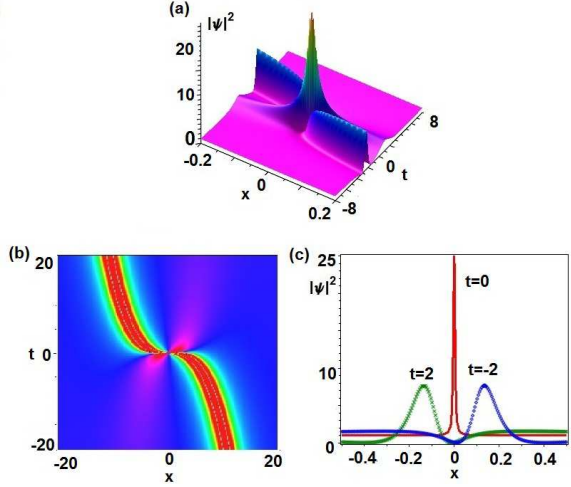}
\caption{(a) The interaction of two solitons (\protect\ref{soliton2}); (b)
2D intensity contour plots; and (c) cross sections for two-solitons (\protect
\ref{soliton2}) for $\protect\alpha =-5,\,\protect\gamma =-1/12$, and $%
\protect\delta =1$.}
\label{fig2-2soliton}
\end{figure}

Some profiles of two-soliton solutions (\ref{soliton2}) are displayed in
Figs.~\ref{fig-2soliton} and \ref{fig2-2soliton} for $\alpha =1,\,\delta =1$
and $\alpha =-5,\,\delta =1$, which display elastic interactions of two
solitary waves. It follows from Fig.~\ref{fig2-2soliton} that the amplitude
of one solitary wave grows to a high limit value, and for the other solitary
wave the amplitude falls to a low limit value at $|x|\rightarrow \infty $.

\section{Conclusions and discussions}

We have explicitly obtained the first- and second-order RW (rogue-wave) and
rational-soliton solutions of the integrable fifth-order NLS equation
belonging to the NLS hierarchy. It includes the fifth-order-dispersion and
quintic-nonlinearity terms with three arbitrary real parameters [$\alpha
,\,\gamma $, and $\delta $ in Eq. (\ref{qnls})]. Several familiar integrable
models, such as the classical-NLS, Hirota, and LPD equations are particular
cases of the FONLS equation, corresponding to particular values of $\alpha
,\,\gamma ,$ and $\delta $, therefore, the previously studied RW solutions
of those equations can be obtained as particular cases of our solutions (\ref%
{rog1sol}) and (\ref{rog2sol}).

The MI (modulation instability) of the flat CW background has been
investigated in the framework of the FONLS equation, as it drives the
formation of the RWs. The first-order RW exhibits one peak and two mutually
symmetric depressions, while the second-order pattern generates an extra
peak-depression pair, in a limited interval of the evolution variable. The
location and motion of the peaks and depressions have been identified, in
analytical and numerical forms. The structure and evolution of the RWs were
elucidated by means of the density plots. Moreover, for the special case $%
\gamma =-1/12$, we obtain the corresponding one- and two-soliton solutions
of Eq.~(\ref{qnls}), which are rational and W-shaped, being different from
the usual bright and dark solitons.

We expect that the predicted results may be observed experimentally in
various physical settings described by the general FONLS equation. It may be
interesting to extend the analysis by further producing more complex
patterns, such as those corresponding to third-order RWs and rational
solitons, which will be studied in the future.

\section*{Acknowledgements}

This work was supported by the National Natural Science Foundation of China
(grants Nos. 61178091,\thinspace 11326165,\thinspace 11202178), and by the
Zhejiang Provincial Natural Science Foundation of China under Grant No.
LQ12A01008.

\begin{widetext}

\section*{Appendix A}

The explicit form of the system of equations $\{\partial\left(|\psi _{1}(x,t)|^{2}\right)/\partial t=0,\, \partial\left(|\psi _{1}(x,t)|^{2}\right)/\partial x=0\}$:
\baselineskip=15pt
\bee \nonumber\begin{array}{l}
(t+6\,x\alpha+30\,x\delta)( 3-4\,{t}^{2}-48\,tx \alpha-240\,tx\delta -144\,{x}^{2}{\alpha}^{2}
-1440\,{x}^{2}\alpha\,\delta-3600\,{x}^{2}{\delta}^{2} \vspace{0.1in} \\ \qquad +1728\,{x}^{2}{\gamma}^{2}+288\,{x}^{2}\gamma+12\,{x}^{2})=0, \vspace{0.1in}\\
-64\,x+276480\,{t}^{2}\alpha\,x\delta+2488320\,t{\alpha}^{2}{x}^{2}\delta
+12441600\,t\alpha\,{x}^{2}{\delta}^{2}+663552\,t\alpha\,{x}^{2}{\gamma}^{2} \vspace{0.1in}\\
\qquad  +110592\,t\alpha\,{x}^{2}\gamma+3317760\,t\delta\,{x}^{2}{\gamma}^{2}
+552960\,t\delta\,{x}^{2}\gamma-256\,{x}^{3}-5308416\,{x}^{3}{\gamma}^{4} \vspace{0.1in}\\
\qquad  -1769472\,{x}^{3}{\gamma}^{3}-221184\,{x}^{3}{\gamma}^{2}-12288\,{x}^{3}\gamma
-1152\,t\alpha-5760\,t\delta-6912\,x{\alpha}^{2}-172800\,x{\delta}^{2}  \vspace{0.1in}\\
\qquad  +1536\,{t}^{3}\alpha+7680\,{t}^{3}\delta+331776\,{x}^{3}{\alpha}^{4}
+207360000\,{x}^{3}{\delta}^{4}+768\,x{t}^{2}-9216\,x{\gamma}^{2}-1536\,x\gamma \vspace{0.1in}\\
\qquad  -69120\,x\alpha\,\delta+4608\,t\alpha\,{x}^{2}+23040\,t\delta\,{x}^{2}
+691200\,{t}^{2}{\delta}^{2}x+18432\,x\gamma\,{t}^{2}+110592\,x{\gamma}^{2}{t}^{2} \vspace{0.1in}\\
\qquad  +49766400\,{x}^{3}{\alpha}^{2}{\delta}^{2}+165888\,t{\alpha}^{3}{x}^{2}+165888000\,{x}^{3}
\alpha\,{\delta}^{3}+20736000\,t{\delta}^{3}{x}^{2}\vspace{0.1in}\\
\qquad  +27648\,{t}^{2}{\alpha}^{2}x+6635520\,{x}^{3}{\alpha}^{3}\delta=0.
\end{array}
\ene

\section*{Appendix B}

Functions $D_{2}$, $G_{2}$, and $K_{2}$, which are parts of analytical
solution (\ref{rog2sol}), are given by the following expressions:

\bee  \nonumber\begin{array}{rl}
D_{2}(x,t)= &64t^{6}+384vxt^{5}+192(B^{2}+5v^{2})x^{2}t^{4}
+48t^{4}+256v(3B^{2}+5v^{2})x^{3}t^{3}-64(120\delta +v)xt^{3}\vspace{0.1in} \\
 & \quad +192(B^{2}+5v^{2})(B^{2}+v^{2})x^{4}t^{2}+108t^{2}+9+4(307B^{2}+57600\delta ^{2}+5280\delta v \vspace{0.1in}\\
 &\quad +139v^{2}-272B +64)x^{2}+96(8B-11B^{2}-240\delta v-5v^{2})x^{2}t^{2}+384v(B^{2}+v^{2})^{2}x^{5}t \vspace{0.1in}\\
 & \quad +24(240\delta +17v)xt+192(120B^{2}\delta -7B^{2}v-120\delta v^{2}-3v^{3}+8Bv)x^{3}t +64(B^{2}
 \vspace{0.1in}\\
 &\quad +v^{2})^{3}x^{6}+16(43B^{2}+1440B^{2}\delta v-18B^{2}v^{2}-480\delta v^{3}-13v^{4}-16B^{3}
 +48Bv^{2})x^{4},%
\end{array}
\ene
\bee \nonumber\begin{array}{rl}
G_{2}(x,t)= &-192t^{4}-1152(B^{2}+v^{2})x^{2}t^{2}-768vxt^{3}-288t^{2}-768v(3B^{2}+v^{2})x^{3}t-192(120\delta+7v)xt
 \vspace{0.1in} \\
 &\quad -192(5B^{2}+v^{2})(B^{2}+v^{2})x^{4}-(17B^{2}+240\delta v+11v^{2}-8B)x^{2}+36,%
\end{array}
\ene
\bee \nonumber\begin{array}{rl}
K_{2}(x,t)= & -384xt^{4}-768B(B^{2}+3v^{2})x^{3}t^{2}+192(7B-4)xt^{2}-1536Bv(B^{2}+v^{2})x^{4}t-1536Bvx^{2}t^{3}
 \vspace{0.1in} \\
&\quad -384(120B\delta -3Bv+4v)x^{2}t-192(5B^{3}+240B\delta v+Bv^{2}-4B^{2}+4v^{2})x^{3}\vspace{0.1in} \\
& \quad -384B(B^{2}+v^{2})^{2}x^{5}+24(23B-8).%
\end{array}
\ene

\section*{Appendix C}

Functions $D_{2\gamma}(x,t)$, $G_{2\gamma}(x,t)$, and $K_{2\gamma}(x,t)$, which are parts of analytical
solution (\ref{soliton2}), are given by the following expressions:

\bee \nonumber\begin{array}{rl}
D_{2\gamma}(x,t)= &64\,{t}^{6}+384\,xv{t}^{5}+960\,{v}^{2}{t}^{4}{x}^{2}+48\,{t}^{4}+1280\,{v}^{3}{t}^{3}{x}^{3}
-\left(7680\,\delta+64\,v \right) {t}^{3}x \vspace{0.1in} \\
& +960\,{v}^{4}{t}^{2}{x}^{4}- \left( 23040\,\delta\,v+480\,{v}^{2}\right)
{t}^{2}{x}^{2}+108\,{t}^{2}+384\,{v}^{5}t{x}^{5} \vspace{0.1in} \\
& -\left(23040\,\delta\,{v}^{2}+576\,{v}^{3} \right) t{x}^{3}+ \left( 5760\,
\delta+408\,v \right) tx+64\,{v}^{6}{x}^{6}\vspace{0.1in} \\
&-\left( 7680\,\delta\,{v}^{3}+208\,{v}^{4} \right) {x}^{4}
+ \left( 256+230400\,{\delta}^{2}+
21120\,\delta\,v+556\,{v}^{2} \right) {x}^{2}+9,
\end{array}
\ene
\bee \nonumber\begin{array}{rl}
G_{2\gamma}(x,t)= &36-192\,{t}^{4}-768\,{t}^{3}vx-1152\,{v}^{2}{t}^{2}{x}^{2}-288\,{t}^{2}-768\,{v}^{3}t{x}^{3}
-23040\,tx\delta \qquad \vspace{0.1in} \\
&-1344\,xvt-192\,{v}^{4}{x}^{4}-23040\,{x}^{2}\delta\,v-1056\,{x}^{2}{v}^{2},
\end{array}
\ene
\bee \nonumber\begin{array}{rl}
K_{2\gamma}(x,t)=&-192\,x \left[ 4 (t+vt)^2+1 \right].\qquad\qquad\qquad\qquad\qquad\qquad\qquad\qquad\qquad\qquad\qquad
\end{array}
\ene

\end{widetext}

\end{document}